\documentclass[12pt]{article}
\usepackage{graphicx}

\newsavebox{\sboxpubnumber}
\newsavebox{\sboxpubdate}
\newcommand{\pubdate}[1]{\begin{lrbox}{\sboxpubdate}{#1}\end{lrbox}}
\newcommand{\pubnumber}[1]{\begin{lrbox}{\sboxpubnumber}{\begin{tabular}{l} #1 \\
				 \usebox{\sboxpubdate}
				 \end{tabular}}
                           \end{lrbox}
                           \pubblock}
\newcommand{\Title}[1]{\begin{center} {\Large #1 } \end{center}}
\newcommand{\Author}[1]{\begin{center}{ \sc #1} \end{center}}
\newcommand{\Address}[1]{\begin{center}{ \it #1} \end{center}}

\newcommand{\pubblock}{\rightline{
			\usebox{\sboxpubnumber}}}
\newenvironment{Abstract}{\begin{quotation}  }{\end{quotation}}
\newenvironment{Presented}{\begin{quotation} \begin{center}
             PRESENTED AT\end{center}\bigskip
      \begin{center}\begin{large}}{\end{large}\end{center}
      \end{quotation}}
\newcommand{\Acknowledgements}{\bigskip  \bigskip \begin{center} \begin{large}
             \bf ACKNOWLEDGEMENTS \end{large}\end{center}}

\newcommand{\beq}{\begin{equation}}
\newcommand{\eeq}{\end{equation}}
\newcommand{\beqa}{\begin{eqnarray}}
\newcommand{\eeqa}{\end{eqnarray}}

\begin{document}

\begin{titlepage}
\pubdate{\today}                            
\pubnumber{HD-TH-01-48, CERN-TH/2002-010, NORDITA-2002-3 HE}  

\vfill
\Title{Some aspects of collisional sources for electroweak baryogenesis}
\vfill
\Author{Kimmo Kainulainen \footnote{University of Jyv\"askyl\"a, Finland, 
                                    on leave of absence.}}
 
\Address{Theory Division, CERN, CH-1211, Geneva 23, Switzerland and \\
         NORDITA, Blegdamsvej 17, DK-2100, Copenhagen \O , Denmark}
\vfill
\vfill
\Author{Tomislav Prokopec,\footnote{Speaker}
            Michael G. Schmidt, and Steffen Weinstock}
\Address{Institut f\"ur Theoretische Physik, Universit\"at Heidelberg \\
          Philosophenweg 16, D-69120 Heidelberg, Deutschland/Germany}
\vfill
\begin{Abstract}
We consider the dynamics of fermions with a spatially varying mass which 
couple to bosons through a Yukawa interaction term and perform a consistent
weak coupling truncation of the relevant kinetic equations. We then use
a gradient expansion and derive the CP-violating source in the collision 
term for fermions which appears at first order in gradients. The collisional 
sources together with the semiclassical force constitute the CP-violating
sources relevant for baryogenesis at the electroweak scale. 
We discuss also 
the absence of sources at first order in gradients
in the scalar equation, and the limitations of the relaxation time 
approximation.
\end{Abstract}
\vfill
\begin{Presented}
    COSMO-01 \\
    Rovaniemi, Finland, \\
    August 29 -- September 4, 2001
\end{Presented}   
\vfill
\end{titlepage}
\def\thefootnote{\fnsymbol{footnote}}
\setcounter{footnote}{0}

\section{Introduction}

The main unsolved problem of electroweak 
baryogenesis~\cite{KuzminRubakovShaposhnikov:1985}
is a systematic computation of the relevant sources in
transport equations. We shall now present a method for controlled derivation
of leading CP-violating sources appearing as a consequence of
collisions of chiral fermions with scalar particles in presence of a scalar field
condensate. We assume the following picture~\cite{CohenKaplanNelson:1991}
of baryogenesis at
a first order electroweak phase transition: when the Universe supercools, 
the bubbles of the Higgs phase nucleate
and grow into the sea of the hot phase. For species that couple to the Higgs
condensate in a CP-violating manner that CP-violating
currents are created at the phase boundary (bubble wall). These currents
then bias baryon number violating interactions mostly in the hot (symmetric) phase,
where the B-violating processes are unsuppressed.
The baryons then diffuse to the Higgs phase, where the B-violating interactions
are suppressed, resulting in baryogenesis. 
 
Kimmo Kainulainen~\cite{Kimmo} has explained
how to systematically derive the CP-violating source in the flow term
of the kinetic equation for fermions. For details see 
{\rm Paper~I}~\cite{KPSW1}.
The source is universal in that its form is 
independent on interactions. It can be represented as the semiclassical force
originally introduced for baryogenesis
in two-Higgs doublet models in~\cite{JoyceProkopecTurok:1994},
and subsequently adapted to the Minimal Supersymmetric
Standard Model (MSSM) in~\cite{ClineJoyceKainulainen:2000}. 
This problem involves computation of CP-violating sources from
charginos, which couple to the Higgs condensate in a manner that
involves fermionic mixing. Here we show how to compute the CP-violating
source in the collision term that arises at first order in gradients.
We work in the simple model of chiral fermions coupled to
a complex scalar field {\it via} the Yukawa interaction with the Lagrangian
of the form~\cite{KPSW1,KPSW2}  
\begin{equation}
     {\cal L} = i\bar{\psi}{\mathbin{\partial\mkern-10.5mu\big/}}
\psi - \bar{\psi}_Lm\psi_R
           -  \bar{\psi}_Rm^*\psi_L  + {\cal L}_{\rm yu},
\label{lagrangian0}
\end{equation}
where ${\cal L}_{\rm yu}$ denotes the Yukawa interaction term
\begin{equation}
{\cal L}_{\rm yu} = - y\phi \bar{\psi}_L\psi_R
           - y \phi^* \bar{\psi}_R\psi_L,
\label{lagrangian1}
\end{equation}
and $m$ is a complex, spatially varying mass term
\begin{equation}
     m(u) \equiv y' \Phi_0 = m_R(u) + i m_I(u) = |m(u)|\mbox{e}^{i\theta(u)}.
\label{mass1}
\end{equation}
Such a mass term arises naturally from an interaction with a scalar 
field condensate $\Phi_0 = \langle \hat \Phi\rangle$. This 
situation is realised for example by the Higgs field condensate of a first
order electroweak phase transition in supersymmetric models.
When $\phi$ in~(\ref{lagrangian1}) is the Higgs field
the coupling constants $y$ and $y'$ are identical;
our considerations are however not limited to this case.

The dynamics of quantum fields can be studied by considering the equations 
of motion arising from the two-particle irreducible effective action 
(2PI)~\cite{CornwallJackiwTomboulis:1974}
in the Schwinger-Keldysh closed-time-path 
formalism~\cite{Schwinger:1961,ChouSuHaoYu:1985}.
This formalism is for example appropriate for studying thermalization 
in quantum field theory \cite{BergesCox:2000}.

\begin{figure}[htb]
    \centering
    \includegraphics[height=0.6in,width=4.in]{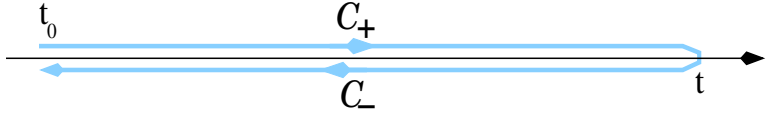}
    \caption{
The Schwinger {\it closed-time-path} (CTP) used in the derivation
of the 2PI effective action~(\ref{EffectiveAction}). 
}
\label{fig:ctp}
\end{figure}
We are interested in the dynamics of the fermionic and bosonic two-point
functions 
\beqa
S_{\alpha\beta}(u,v) 
 &=& -i\left<T_{{\cal C}}\big[\psi_\alpha(u)\bar{\psi}_\beta(v)\big]\right>
\label{S}
\\
\Delta(u,v) 
 &=& -i\left<T_{{\cal C}}\big[\phi(u){\phi}^\dagger(v)\big]\right>
\label{Delta}
\eeqa
where the time ordering $T_{{\cal C}}$ is along the Schwinger
contour shown in figure~\ref{fig:ctp}, which is suitable for the dynamics
of out-of-equilibrium quantum fields.

\section{Effective action and self-energies}

\begin{figure}[htb]
    \centering
    \includegraphics[height=1.in,width=4.in]{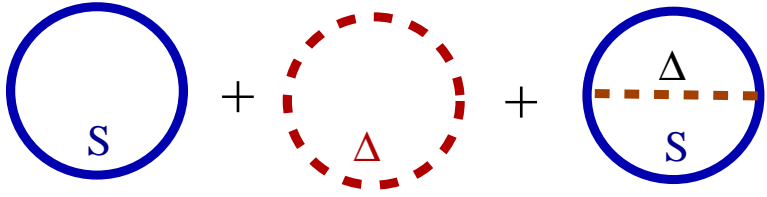}
    \caption{
\small
The diagrams contributing to the 2PI effective 
action~(\ref{EffectiveAction0}-\ref{EffectiveAction2})
up to two loops for the Lagrangian~(\ref{lagrangian0}) with the
fermion-scalar Yukawa coupling term~(\ref{lagrangian1}). 
The full (dressed) fermionic and bosonic
propagators are denoted by $S$ ({\it solid blue lines}) and 
$\Delta$ ({\it dashed red lines}), respectively.
}
\label{fig:2pi}
\end{figure}

The 2PI effective action can be in general written in the 
form~\cite{CornwallJackiwTomboulis:1974,ChouSuHaoYu:1985} 
\beq
\Gamma_{\rm eff} = \Gamma_0 +  \Gamma_1 + \Gamma_{\rm 2PI},
\label{EffectiveAction}
\eeq
where the tree-level and one-loop actions read
\beqa
\Gamma_0 &=& i\int_{{\cal C}} d^4 u\, d^4 v \big[\Delta_0^{-1}(u,v) \Delta (v,u)
   -S_0^{-1}(u,v) S (v,u)\big]
\label{EffectiveAction0}\\
\Gamma_1 &=&
  i \int_{{\cal C}} d^4 u  \big[\ln  \Delta^{-1} (u,u)-\ln S^{-1} (u,u)\big],
\label{EffectiveAction1}
\eeqa
and $\Delta_0^{-1}$ and $S_0^{-1}$ are the tree-level {\it inverse} propagators:
\beqa
S_0^{-1}(u,v) &=& (i \partial\!\!\!/_u - m(u) P_R - m^*(u) P_L) 
   \delta_{{\cal C}}(u-v)
\label{S0}
\\
\Delta_0^{-1}(u,v) &=& (-\partial^2_u - m_\phi^2(u)) \delta_{{\cal C}}(u-v).
\label{Delta0}
\eeqa
Here $P_{L,R} = (1\mp\gamma^5)/2$, and 
$\delta_{{\cal C}}$ is the $\delta$-function along the contour $\cal C$.
$\Gamma_{2PI}$ in~(\ref{EffectiveAction}) contains higher order quantum
corrections, which can be for example studied in the loop expansion. 
At two loops $\Gamma_{2PI}$ reads
\beq
\Gamma_{\rm 2PI} \rightarrow \Gamma_{2}[S,\Delta]
  =   -y^2 \int_{{\cal C}} d^4 u\, d^4 v
      \;{\rm Tr}\big[P_RS(u,v)P_LS(v,u)\big]\, \Delta(u,v)
,
\label{EffectiveAction2}
\eeq
which can be easily computed from the two-loop diagram in 
figure~\ref{fig:2pi}. The one-loop self-energies are then obtained
from~(\ref{EffectiveAction2}) by taking the functional derivatives 
\beqa
\Sigma'(u,v) &=& - i \frac{\delta\Gamma_{\rm 2PI}}{\delta S (v,u)}
  = i y^2\, \big[
    P_L S(u,v) P_R \Delta(v,u) + P_R S(u,v)P_L \Delta(u,v)
\big]
\label{Sigma}
\\
\Pi'(u,v) &=& i \frac{\delta\Gamma_{\rm 2PI}}{\delta \Delta(v,u)}
  = - i y^2\, {\rm Tr}\,\big[P_L S(v,u) P_R S(u,v)\big]
.
\label{Pi}
\eeqa
\begin{figure}[htb]
    \centering
    \includegraphics[height=2.2in,width=5.in]{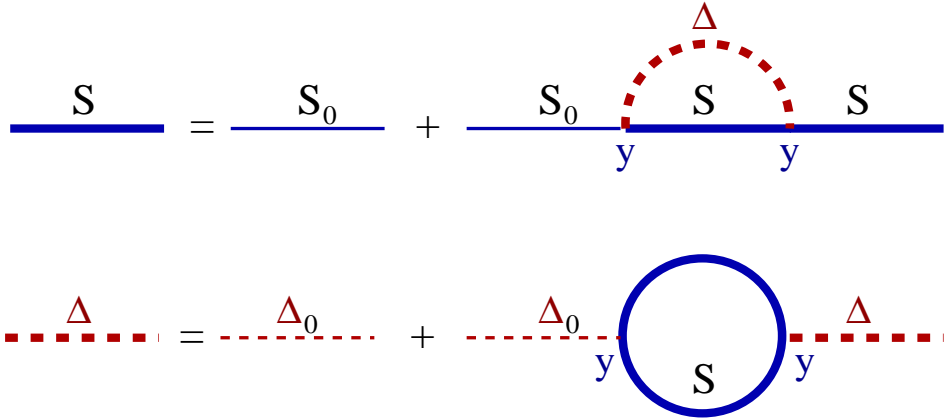}
    \caption{
\small
The Dyson-Schwinger equations at one loop obtained by varying 
the 2PI effective action~(\ref{EffectiveAction0}-\ref{EffectiveAction2}) 
with respect to the fermionic and bosonic propagators. The corresponding
tree-level Lagrangian is given by~(\ref{lagrangian0}-\ref{lagrangian1}). 
The full (dressed) fermionic and bosonic
propagators are denoted by $S$ ({\it solid blue lines}) and 
$\Delta$ ({\it dashed red lines}), respectively.
}
\label{fig:dse}
\end{figure}

The equation of motion for $S$ and $\Delta$ are obtained by varying 
the effective action with respect to $S$ and $\Delta$. 
The resulting equations are simply
\beqa
S_0^{-1}\otimes S &=& \delta_{{\cal C}} + \Sigma'\otimes S
\label{DSeomS}
\\
\Delta_0^{-1}\otimes\Delta &=& \delta_{{\cal C}} + \Pi'\otimes \Delta
,
\label{DSeomDelta}
\eeqa
%
where $\otimes$ denotes a convolution with respect to contour integration. 
These are the Dyson-Schwinger integro-differential equations  
diagrammatically shown in figure~\ref{fig:dse}.
The simple look of these equations is deceptive, since they involve 
integration over the closed-time-path in figure~\ref{fig:ctp}.
To proceed we use the Keldysh reformulation of the problem, according to 
which the contour ${\cal C}$ is split into two parts: $(t_0,t)$ in the positive
time direction and $(t,t_0)$ in the negative time direction; finally
we set $t_0\rightarrow -\infty$. This corresponds to the replacements
\beqa
\int_{{\cal C}} d^4u &\longrightarrow& 
  \sum_{a=\pm 1}a\int_{t_0\rightarrow -\infty}^t d^4u 
\nonumber\\
S(u,v) &\longrightarrow& S^{ab}(u,v)
\nonumber\\
\Delta(u,v) &\longrightarrow& \Delta^{ab}(u,v)
\nonumber\\
\delta_{{\cal C}}(u-v)  &\longrightarrow& a\delta_{ab}\delta(u-v)
\label{ctp-t}
\eeqa
in the effective action~(\ref{EffectiveAction}-\ref{EffectiveAction2}).
This procedure naturally leads to the Keldysh $2\times2$ formulation for 
the two-point functions~(\ref{S}-\ref{Delta}) in which the off-diagonal 
elements of the fermionic Wigner functions correspond to
\beqa
S^{<}(u,v) &\equiv& S^{+-}(u,v) = i\langle \bar\psi(v) \psi(u) \rangle
\nonumber\\
S^{>}(u,v) &\equiv& S^{-+}(u,v) = - i\langle  \psi(u)\bar\psi(v)\rangle,
\label{S-ab}
\eeqa
and similarly the bosonic ones~(\ref{Delta}) are
\beqa
\Delta^{<}(u,v) &\equiv& \Delta^{+-}(u,v) 
             = - i\langle \phi^\dagger(v) \phi(u) \rangle
\nonumber\\
\Delta^{>}(u,v) &\equiv& \Delta^{-+}(u,v) 
             = - i\langle  \phi(u)\phi^\dagger(v)\rangle.
\label{Delta-ab}
\eeqa
%
In the Keldysh representation it is convenient to redefine the
self-energies~(\ref{Sigma}-\ref{Pi}) as
\beqa
\Sigma^{ac}(u,v) &\equiv &  a\Sigma'^{ac}(u,v) c
\label{Sigma-ac}
\\
\Pi^{ac}(u,v) &\equiv&  a\Pi'^{ac}(u,v) c .
\label{Pi-ac}
\eeqa

\section{Kinetic equations}

 When written explicitly the kinetic equations for fermions~(\ref{DSeomS})
and bosons~(\ref{DSeomDelta}) become
\beqa
(\frac i2\partial\!\!\!/+ k\!\!\! / - \hat mP_R - \hat m^*P_L) S^{<}
  - e^{-i\diamond}\{ \Sigma_R, S^{<}\}
  - e^{-i\diamond}\{ \Sigma^{<}, S_R\}
    &=& {\cal C}_\psi 
\label{Seom-<w}
\\
(-\frac 14 \partial^2 + k^2  + ik\cdot\partial - \hat m^2) \Delta^{<}
  - e^{-i\diamond}\{ \Pi_R, \Delta^{<}\}
  - e^{-i\diamond}\{ \Pi^{<}, \Delta_R\}
    &=& {\cal C}_\phi ,
\label{Deltaeom-<w}
\eeqa
where 
\beqa
    S^<(x,k) &=& \int d^{\,4} r \, e^{ik\cdot r} S^<(x+r/2, x-r/2)
\label{wignerS}
\eeqa
defines the Wigner functions for fermions, and there is a similar expression for
bosons. $\Pi_R$, $\Sigma_R$, $S_R$ and $\Delta_R$ denote the hermitean parts of 
the self-energies and two-point functions, respectively. We postpone a discussion
of the physical relevance of the contributions involving the hermitean parts 
to a later publication. Eq.~(\ref{wignerS}) defines
a Wigner transform, which is useful to separate the dependence
on the slowly varying average ({\it macroscopic}) coordinate $x = (u+v)/2$
from that on the {\it microscopic} coordinate $r = u-v$. 

The bilinear operator $\diamond$ in Eqs.~(\ref{Seom-<w}-\ref{Deltaeom-<w}) 
is defined as 
\beq
\diamond\{A,B\} \equiv 
    \frac 12 \Big(\partial_x A\cdot\partial_k B
                  - \partial_k A\cdot\partial_x B \Big),
\label{diamond}
\eeq
the mass terms read
\beqa
\hat m &=& m(x) e^{-\frac i2 
{\stackrel{\leftarrow}{\partial}}
\cdot\;\partial_k}
\label{m-hat}
\\
\hat m_\phi^2 &=& m_\phi^2(x) 
    e^{-\frac i2 \stackrel{\leftarrow}{\partial}\cdot\;\partial_k}
,
\label{m-hat-phi}
\eeqa
and the collision terms are of the form
\beqa
 {\cal C}_\psi  &=&
 - \frac 12 e^{-i\diamond}\Big( \{\Sigma^>, S^<\} - \{\Sigma^<, S^>\} \Big),
\label{C-psi}
\\
 {\cal C}_\phi  &=& 
-\frac 12 e^{-i\diamond}\Big( \{\Pi^>, \Delta^<\} - \{\Pi^<, \Delta^>\} \Big)
.
\label{C-phi}
\eeqa
The one-loop expressions for the self-energies can be inferred from
Eqs.~(\ref{Sigma-ac}-\ref{Pi-ac}) and~(\ref{Sigma}-\ref{Pi}):
\beqa
 \Sigma^{<,>}(k,x) &\equiv& \Sigma^{+-,-+}(k,x) 
\nonumber\\
&=& i y^2\int \frac{d^4k'd^4k''}{(2\pi)^8}\,\big[
    (2\pi)^4\delta(k-k'+k'') P_L S^{<,>}(k',x) P_R \Delta^{>,<}(k'',x) 
\nonumber\\
&& \qquad\qquad \qquad 
  + (2\pi)^4\delta(k-k'-k'') P_R S^{<,>}(k',x) P_L \Delta^{<,>}(k'',x)
\big]
\label{Sigma-<>w}
\\
 \Pi^{<,>}(k,x) &\equiv& \Pi^{+-,-+}(k,x)  
\nonumber\\
 &=&\!\!\!
 - i y^2\!\! \int \frac{d^4k'd^4k''}{(2\pi)^8} (2\pi)^4\delta(k+k'-k'')
  {\rm Tr}\,\big[P_R S^{>,<}(k',x) P_L S^{<,>}(k'',x)\big]
.\quad
\label{Pi-<>w}
\eeqa

\section{Wigner functions}

 We are interested in modeling the dynamics of fermions and bosons 
in the presence of a Higgs field condensate 
of growing bubbles at a first order electroweak phase transition.
When the bubble walls are thick
we can expand in gradients of the condensate.
This expansion is accurate for quasiparticles whose 
de Broglie wavelength $\ell_{dB}$ is small when compared to the scale of variation 
of the background, which is specified by the phase boundary thickness $L_w$.
Since equilibrium considerations yield $L_w\sim 5-15$, and the de Broglie wavelength
is typically given by the inverse temperature, $\ell_{dB}\sim 1/T$, we have
$\ell_{dB}\partial_x \sim \ell_{dB}/L_w\ll 1$, so that the expansion in gradients 
is justified.

The problem can be further simplified by noting that typically large bubbles
are almost planar and slow, such that it suffices to
keep the leading order terms in the bubble wall velocity, and expand to leading 
nontrivial order in gradients. So when written in the 
rest frame of the bubble wall (wall frame),
Eqs.~(\ref{m-hat}-\ref{m-hat-phi}) simplify to
$\hat m(z) = m + \frac i2  m'\partial_{k_z} + ..$ and 
$\hat m_\phi^2(z) = m_\phi^2 + \frac i2 {m_\phi^2}' \partial_{k_z} + ..\,$.

In Paper~I we have shown that the fermionic Wigner function~(\ref{S-ab}) 
acquires a nontrivial contribution at first order in gradients. 
In the presence of a scalar condensate, when one neglects particle interactions,
a wall moving in $z$-direction conserves spin in $z$-direction. 
This implies that one can write ({\it cf.\ } Paper~II) 
the Wigner function in the following spin-diagonal form 
\beq
S^{<,>} = \sum_{s=\pm} S_s^{<,>},
\label{S<>}
\eeq
where 
\beq
S_s^{<,>} = iP_s \Bigl[s \gamma^3\gamma^5 g_0^{s<,>}-\gamma^3 s g_3^{s<,>}
    + g_1^{s<,>} - i\gamma^5 g_2^{s<,>} \Bigr].
\label{S-1}
\eeq
Here $P_s$ denotes the spin projector
\beqa
P_s &=& \frac 12 (1+ sS_z), 
\label{Ps}
\eeqa
where
\beqa
 S_z &=& 
\frac{1}{\tilde k_0}(k_0\gamma^0 - \vec k_\|\cdot\vec \gamma_\|) \,\gamma^3\gamma^5
\label{S_z}
\eeqa
is the spin operator in $z$-direction, and  
$\tilde {k}_0 = {\rm sign}[k_0]\sqrt{k_0^2-\vec k_\|^2}$.

 The quantity $g_0^{s<,>}$ in Eq.~(\ref{S-1}) is a measure for
the phase space density of particles with spin $s$. When written in the wall frame
and to first order in gradients, the solution for $g_0^{s<,>}$ has the form 
\beqa
 g_0^{s<} &=& 
         2\pi\delta\Big(k^2 - |m|^2 + \frac{1}{\tilde k_0}s|m|^2\theta'\Big) 
              |\tilde {k_0}| n(k,x)
\nonumber\\
 g_0^{s>} &=& 
     - 2\pi\delta\Big(k^2 - |m|^2+ \frac{1}{\tilde k_0}s|m|^2\theta'\Big)
              |\tilde {k_0}| (1 - n(k,x)).
\label{g0<>1}
\eeqa
In thermal equilibrium we have 
\beq
n(k,x) \rightarrow n_0 = \frac{1}{e^{\hat k_0/T}+1},\qquad 
   \hat k_0 = \gamma_w (k_0 + v_w k_z),
\label{n0}
\eeq
where $\vec v_w = v_w \hat z$ is the wall velocity
and $\gamma_w = 1/\sqrt{1-v_w^2/c^2}$. 
The quantities $g_i^{s<,>}$ ($i=1,2,3$) in~(\ref{S-1}) can be related
to $g_0^{s<,>}$ by (see Paper I \& II)
\beqa
g_1^{s<,>} &=& \frac{1}{\tilde k_0} \bigg[
 m_R g_0^{s<,>} - \frac {s}{2\tilde k_0}\partial_z (m_I g_0^{s<,>})
                - \frac {sm_I'}{2\tilde k_0}\partial_{k_z} (k_z g_0^{s<,>})
\bigg]
\nonumber\\
g_2^{s<,>} &=& \frac{1}{\tilde k_0} \bigg[
 m_I g_0^{s<,>} + \frac {s}{2\tilde k_0}\partial_z (m_R g_0^{s<,>})
                + \frac {sm_R'}{2\tilde k_0}\partial_{k_z} (k_z g_0^{s<,>})
\bigg]
\nonumber\\
g_3^{s<,>} &=& \frac{1}{\tilde k_0} \bigg[
 sk_z g_0^{s<,>} + \frac {|m|^2\theta'}{2\tilde k_0}\partial_{k_z} g_0^{s<,>}
\bigg].
\label{ce-S}
\eeqa
It is now easy to check that, to leading order in gradients,
the Kubo-Martin-Schwinger (KMS) condition
\beq
S_0^> = - e^{\hat k_0/T}S_0^<
\label{KMS-S0}
\eeq
holds when $n=n_0$. The first order solution~(\ref{S-1}) with $n=n_0$
does {\em not} obey the KMS condition however, except in the limit
of a static wall $v_w\rightarrow 0$.  

The solution for the scalar field Wigner function is much simpler.
One can show~\cite{KPSW1} that, to the order $\hbar$ in gradient expansion,
there is no source for bosons, so that the bosonic Wigner function 
in the wall frame can be approximated by 
\beqa
\Delta^< &=& \frac{\pi}{\omega_\phi}\left[\delta(k_0-\omega_\phi)f_\phi
 + \delta(k_0+\omega_\phi)(1+\bar f_\phi)\right]
\nonumber\\
\Delta^> &=& \frac{\pi}{\omega_\phi}\left[\delta(k_0-\omega_\phi)(1+f_\phi)
 + \delta(k_0+\omega_\phi)\bar f_\phi\right]
\label{Delta<>}
\eeqa
where $f_\phi(k_z)$ and $\bar f_\phi(-k_z)$ denote the occupation number of 
particles and antiparticles, respectively, such that in thermal equilibrium 
we get the standard Bose distribution function
\beq
f_{\phi 0} = \frac{1}{e^{\gamma_w(\omega_\phi+v_w k_z)/T}-1}
,\qquad
\bar {f}_{\phi 0} = \frac{1}{e^{\gamma_w(\omega_\phi-v_w k_z)/T}-1}
\label{Bose}
\eeq
with $\omega_\phi=\sqrt{\vec k^2+m_\phi^2}$. Note that in thermal 
equilibrium the KMS condition, $\Delta_0^> = e^{k_0/T}\Delta_0^<$,
is satisfied.

\section{Collisional sources}

\subsection{Collisional source in the scalar equation}

Making use of equations~(\ref{Deltaeom-<w}), (\ref{C-phi}) 
and~(\ref{Pi-<>w}-\ref{ce-S}), one can show~\cite{KPSW2} that to 
first order in gradients the scalar field collision integral 
${\cal C}_{\phi}$ is proportional to delta functions:
%
\beq
{\cal C}_{\phi} \propto 
    \delta(k_0-\omega_\phi)  +  \delta(k_0+\omega_\phi).
\label{scalar Source}
\eeq
Because of the positive relative sign between $f_\phi$ and 
${\bar f}_\phi$ in the scalar Wigner function~(\ref{Delta<>}), 
the source 
arising from ${\cal C}_{\phi}$
gives rise to a CP-even and 
spin independent contribution to the scalar kinetic 
equation~(\ref{Deltaeom-<w}), which is thus of no relevance for 
baryogenesis. As a consequence, there is no CP-violating source 
for {\it stops}~\cite{Riotto:1998} in the collision term at first 
order in gradients. Our analysis is however based on a one-loop 
approximation for the self-energies. To make the study more complete,
it would be desirable
to perform a two-loop calculation of the source.

\subsection{Collisional source in the fermionic equation}

Before proceeding to calculate the source in the fermionic kinetic 
equation, we first prove that {\it there has to be} a nonvanishing 
source in the fermionic collision term.

Since the equilibrium solution~(\ref{S-1}-\ref{n0}) for the 
Wigner function $S^{<,>}$ does not satisfy the KMS condition, 
the collision term does not 
vanish in the equilibrium defined by Ansatz~(\ref{n0}).
We shall now show that it is not possible to avoid this outcome 
by choosing $n$ differently.

First recall that for a static wall with $v_w=0$ the equilibrium 
solution~(\ref{S-1}-\ref{n0}) {\em does} satisfy the KMS condition. 
Now try to modify the equilibrium solution for $S^{<,>}$ 
as follows
\beqa
S^{<} &=& 2i{\cal A}(n_0+\delta n),
\qquad
 S^{>} =  -2i{\cal A}(1-n_0- \delta n )
,
\label{S<>0-generic}
\eeqa
where $\delta n$ is some 
scalar function of first order in gradients.
The spectral function
\beq
{\cal A} = {\cal A}_{0} + {\cal A}_1
\label{spectral-function}
\eeq
contains the spinor structure at leading and first order in derivatives,
respectively, and the $\delta$-function, 
$\delta(k^2-|m|^2 + s|m|^2\theta'/\tilde k_0)$, as specified 
by equations~(\ref{ce-S}). ${\cal A}_1$ contains derivatives
which spoil the KMS condition when acting on $n_0(\hat k_0)$. 
We now seek $\delta n$ such that $S^{>} = -e^{\beta \hat k_0}S^{<}$
be satisfied. Working to first order in gradients one 
arrives at the following condition
\beqa
{\cal A}_0\delta n 
                   &=& n_0 {\cal A}_1 1 - {\cal A}_1n_0
.
\label{delta n0}
\eeqa
where 
${\cal A}_1$ acts as an operator on $1$ and $n_0$, respectively. 
Since ${\cal A}_0$ and  ${\cal A}_1$ have different spinor structure,
there is no scalar function $\delta n$ that solves this equation.
Hence there are indeed collisional sources that cannot be removed by 
a change of thermal equilibrium that is local in momentum.
\begin{figure}[tbp]
    \centering
    \includegraphics[height=3.1in,width=4.7in]{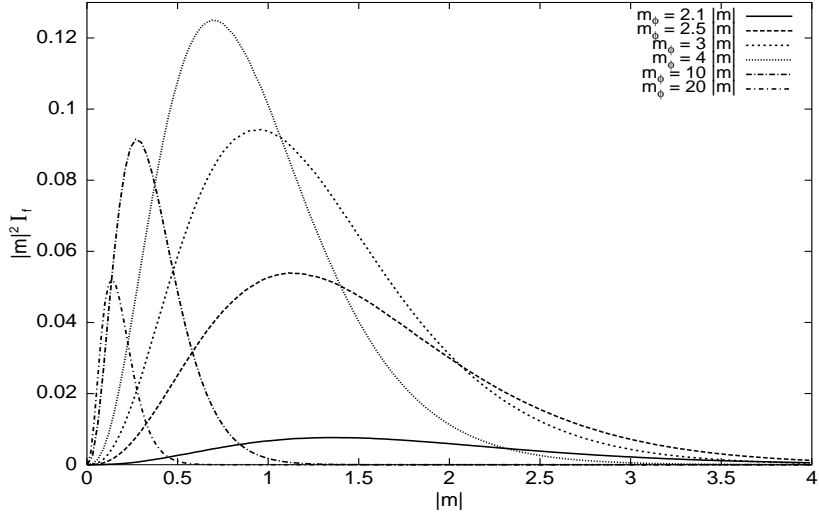}
\vskip -0.3in
\caption{
\small
Shown as a function of $|m|/T$ is the expression 
$|m|^2{\cal I}_f(|m|,m_\phi)$, which appears in the collisional source
(46) arising in the fermionic kinetic equation at one loop. 
For simplicity we have set $T=1$.
}
\label{fig:Cf-source}
\end{figure}

\medskip

In order to obtain the relevant collisional source in the kinetic 
equation for fermions, we need to multiply Eq.~(\ref{C-psi}) 
by $-P_s$, take the real part of the spinorial trace and truncate 
at first order in gradients.

To get the collisional source in the vector current continuity equation
one should integrate the collision term over the momenta. 
It is not hard to see that the integrand is symmetric under 
the change of variables $k_z \rightarrow -k_z$ and 
$k'_z \rightarrow -k'_z$, which then implies that 
\beq
\int_{\pm} \frac{d^4 k}{(2\pi)^4} {\cal C}_{\psi(0)}  = 0,
\label{FermionicSource0}
\eeq
where ${\cal C}_{\psi(0)} = - {\rm Tr}[P_s {\cal C}_{\psi}]$ and 
$\int_{\pm}$ denotes integration over the positive and negative 
frequencies, respectively.
This means that there is no collisional source in the continuity
equation for the  vector current. 

The source in the first velocity moment equation, the Euler fluid 
equation, is however nonvanishing. One can show that, to leading order 
in $v_w$, the source can be written in the following form
\beqa
2\int_{\pm}\frac{d^4k}{(2\pi)^4}\frac{k_z}{\omega_0} {\cal C}_{\psi (0)} 
 &=& \pm  v_w y^2\frac { s|m|^2\theta'}{64\pi^3 T}
{\cal I}_f(|m|,m_\phi)
\label{FermionicSource1}
\eeqa
where the function ${\cal I}_f(|m|,m_\phi)$ 
contains a complicated integral expression.
To get a quantitative estimate of the source, we perform numerical 
integration of ${\cal I}_f$ and plot the result in 
figure~\ref{fig:Cf-source}. Note that for $|m|\gg T$ ($T=1$ in the 
figure) the source is Boltzmann suppressed, while for $|m|\ll T$ it 
behaves as $\propto (|m|/T)^{3/2}$. The source peak shifts towards 
the infrared as $m_\phi/|m|$ increases, and the magnitude drops 
drastically when one approaches the mass threshold $m_\phi = 2|m|$,
as requested for scalar particle decay and inverse decay processes. 

\subsection{Collisional sources in the relaxation time approximation}

The methods used in literature for computation of sources from 
the collision term in scalar and fermionic equations 
\cite{spontaneous,Riotto:1998} can be in many cases rephrased as
the relaxation time approximation:
%
\beq
{\cal C}_{\psi si(0)} \approx -\Gamma_{i} \left(f_{si} - f_{si0}\right),
\label{RelaxationTimeAppox}
\eeq
where $f_{si}$ denotes the true particle density of spin $s$ and flavour $i$, 
$f_{si0}$ is the equilibrium particle density (for a moving wall), and 
$\Gamma_{i}$ is the relevant relaxation rate, which is usually assumed to be 
given by the elastic scattering rate. In Paper~II we have argued that, 
when the equilibrium distribution function is taken to be 
$f_{si0} = 1/(e^{(\omega_{si\pm} + v_w k_z)/T} +1)$, the collision 
term~(\ref{RelaxationTimeAppox}) contains no 
CP-violating source of relevance for baryogenesis. 
When one however takes (incorrectly) for the equilibrium distribution function
$f_{si0} \rightarrow  1/(e^{(\omega_{0} + v_w k_z)/T} +1)$, 
Eq.~(\ref{g0<>1}) implies the following form for the CP-violating contribution
to the collisional source 
\begin{equation}
 - \Gamma_{i}n_{si}^0 \equiv
   - \Gamma_{i}(n_{si+}^0 - n_{si-}^0)
     =  - \Gamma_{i}\frac{s |M_i|^2\Theta_i'}{16\pi^2}\; {\cal J}_0(|M_i|/T),
\label{n0si}
\end{equation}
where $n_{si\pm}^0 \equiv \int_\pm ({d^4k}/{(2\pi)^4})g_{00}^s$ and 
$M_i$ and $\Theta_i$ denote the masses and CP-violating phases of fermions.
The detailed form of the function ${\cal J}_0$ is discussed in Paper~II 
and it is not of importance for the purpose of this talk. 

Apart from parametric dependencies issues, the most worrisome feature about 
the relaxation time approximation interpreted such that it contains
the source~(\ref{n0si}), is that it yields a nonvanishing source 
(\ref{n0si}) in the {\em vector} continuity equation, which we found to
vanish based on equation~(\ref{FermionicSource0}).  Moreover, the 
source~(\ref{n0si}) does not vanish in the static limit  $v_w\rightarrow 0$,
and hence it is clearly unphysical with no apparent relevance for baryogenesis.

We finally note that the CP-violating sources~(\ref{FermionicSource1}), 
(\ref{n0si}) discussed here, when applied to charginos in the MSSM,
show the parametric dependence
$|M_i|^2\Theta_i'\propto(h_1h_2)'$ on the Higgs field
{\it vev}'s $h_1$ and $h_2$. This is to be compared with
Refs.~\cite{CarenaQuirosRiottoViljaWagner,CarenaMorenoQuirosSecoWagner},
where a source proportional to $h_1h_2'-h_2h_1'$ was found and claimed 
to be important for baryogenesis.

\section{Conclusions}

We have studied collisional sources appearing at first order in gradients
of a spatially varying mass for the model of chiral
fermions interacting with a scalar field {\it via} a standard Yukawa interaction. 
This model is relevant for baryogenesis from fermions interacting with 
the Higgs condensate on growing bubbles at a strongly first order electroweak
phase transition. The self-energies have been approximated by one-loop 
expressions. We have argued that, at first order in gradients, there is no 
collisional source in the scalar kinetic equation, indicating that baryogenesis
sourced by scalar particles is highly suppressed. We have then proven
that there is a CP-violating source in the fermionic equation and 
performed a quantitative analysis of the source. This source,
together with the one from the semiclassical force, comprise the relevant
sources for baryogenesis at the electroweak scale. Finally, we have argued  
that the collision term, when modeled in the relaxation time approximation,
contains no CP-violating sources. In order to perform a full quantitative
assessment of collisional sources, a two-loop analysis of self-energies
is required, which is a work in progress.

\Acknowledgements

 We wish to thank Michael Joyce for helpful insights at earlier stages
of our work on collisional sources.

\end{document}